\documentstyle[preprint,aps,prb]{revtex}
\begin{document}
\draft
\title{ Localized-orbital computation of linear and nonlinear
susceptibilities }

\author{ Andrea Dal Corso and Francesco Mauri}\address{
Institut Romand de Recherche Num\'erique en Physique
des Mat\'eriaux (IRRMA)  PHB Ecublens 1015 Lausanne, Switzerland}
 \date{\today} \maketitle

\begin{abstract}
We present a method to compute high-order derivatives of the
total energy which can be used in the framework of density functional theory.
We provide a proof of the $2n+1$ theorem for a
general class of energy functionals in which the orbitals are not
constrained to be orthonormal. Furthermore, by combining this result with
a recently introduced Wannier-like
representation of the electronic orbitals,
we find expressions for the static
linear and nonlinear susceptibilities which are much
simpler than those obtained by standard perturbative expansions.
We test numerically the validity of our approach with a 1D model
Hamiltonian.
\end{abstract}

Perturbative techniques are usually applied to density
functional theory (DFT)\cite{DFT,BGT1}
to study the response properties of materials from first principles.
The evaluation of the second-order derivatives
of the total energy yields phonon spectra,\cite{Giannozzi91}
effective charges,\cite{Giannozzi91,Dalcorso1}
dielectric constants,\cite{Hybertsen87a,Resta}
piezoelectric tensors~\cite{Degironcoli,Dalcorso2} and many other
experimentally measurable quantities.
Likewise
the computation of higher-order derivatives permits
the {\it ab-initio} prediction of properties
such as the Raman tensors,\cite{Baroni2} the second and higher-order
susceptibilities, the nonlinear elastic constants etc..

There are already very sophisticated analytical methods to obtain
the values of the
second-order derivatives, and today it is possible to evaluate these
quantities in systems with many atoms per unit cell.\cite{gonze92}
On the contrary, the evaluation of third or higher-order derivatives
relies mainly on finite differences:
the required derivatives are computed by numerical differentiation
of the second-order derivatives. The cost of the finite differentiation
limits the applicability of the technique to small systems and to short
wavelength perturbations.

Closed-form expressions of the third or
higher-order derivatives, obtained by a straightforward application of
quantum-mechanical perturbation theory, are usually cumbersome.

In the case of the second-order susceptibilities, i.e third order derivatives
of the energy with respect to a uniform electric field, the perturbative
expansion provides a formula which apparently diverges in the static
limit. This divergence can be eliminated as shown in
Ref.~(\onlinecite{aspnes}) in the context of a non self-consistent
electronic structure theory. A specific application of the resulting formula
has been performed using a semi-empirical
tight-binding Hamiltonian.\cite{sipe} In order to extend this
scheme to self-consistent DFT
one has to face rather formidable formal difficulties.
An explicit expression for the second-order susceptibility within DFT
has been obtained using a software package for symbolic
manipulation,\cite{Levine91} and it has been applied by just one research
group due to its complexity.\cite{Levine89}

Alternative analytical expressions for the high-order derivatives of the energy
are provided by the $2n+1$ theorem, well known
in standard perturbation theory for many years~\cite{morse} and
recently rewritten in the language of DFT.\cite{gonze89}
This theorem states that the derivatives of the energy up to
order $2n+1$ can be computed
if the change of the wavefunctions is known up to order $n$.
This approach appears particularly promising to compute high-order
derivatives of the total energy with respect to an atomic displacement but,
in the formulation of Ref.~\onlinecite{gonze89}, it is
of no practical use when the perturbation is an
electric field. In fact the formulas contain
the change of the eigenvalues of the Hamiltonian due to the perturbation,
i.e. a quantity which is ill-defined when the perturbation is an electric field
and the wavefunctions are Bloch states.

Recently new methods have been introduced in DFT to solve the electronic
structure problem, mainly to reduce the number of operations
necessary for the numerical solution.
One of these methods~\cite{Mauri} is based on a Wannier-like
representation of the electronic orbitals which are constrained
to be localized in finite regions of the real space.
The localized states are in general nonorthonormal and are obtained
from a direct minimization of the total energy of the system.
The method is very convenient to study systems with many atoms
since the localization of the wavefunctions allows
the computation of the total energy with a workload proportional
to the number of atoms.
At the same time, the application of this technique to a periodic solid
provides a good approximation for the Wannier
functions which are usually difficult to obtain with other techniques.
In Ref.~\onlinecite{Vanderbilt} it was shown that the center of these
Wannier-like functions yields the correct polarization of the
system,\cite{KSV} and that their localization property
can be conveniently used to study the
behavior of a periodic insulating solid inside a uniform electric field.
This approach allowed the computation of the physical properties of a solid
under a finite electric field.
The derivatives of the energy with respect to the electric field
were computed by means of accurate finite difference
calculations.\cite{Vanderbilt}

In this paper we further extend the approach of Ref.~\onlinecite{Vanderbilt}
and Ref.~\onlinecite{gonze89} and we give a method to compute
{\it analytically} high-order derivatives of the energy.
In particular we give a very general derivation of the $2n+1$ theorem,
which does not require the definition of a specific
energy functional, contrary to what was done in previous work.
We then apply our result to obtain the expressions
for the linear and nonlinear susceptibilities in the
Wannier-like representation of the electronic orbitals.\cite{Mauri,Vanderbilt}
The resulting expression for the second-order
susceptibility is much simpler than the one
obtained by standard perturbation theory\cite{aspnes} because the use of the
$2n+1$ theorem allows us to express this third-order derivative of the
energy only as a function of the first order variation of the
wavefunctions.
Furthermore the use in the $2n+1$ theorem of Wannier-like functions
instead of Bloch eigenfunctions,\cite{gonze89} gives an expression
for the second order susceptibility which is well defined also for the
case when the perturbation is a uniform electric field.

We apply our results to a 1D model Hamiltonian
to test the convergence properties of the proposed algorithm.
We compute analytically the first-, second- and third-order derivatives
of the total energy with respect to a uniform electric field
and we compare the results with those of the finite difference calculations.
The third-order derivative is computed for an arbitrary field,
so that the fourth-order derivative is available as well through
finite differences.
The simplicity of our method makes it very well suited
to compute high-order derivatives of the total energy
for real materials in the framework of DFT.

We start with a general proof of the $2n+1$ theorem valid for
an arbitrary total energy functional $E({\bf w},\lambda)$,
where ${\bf w}$ is a vector whose elements are the
coefficients of all the occupied wavefunctions on a given basis
and $\lambda$ is a
parameter measuring the magnitude of the perturbation.
We restrict ourselves to energy functionals where no explicit
constraints---as for example those of orthonormalization---are imposed on the
components of ${\bf w}$.\cite{Mauri}
For a given $\lambda$ the total energy is defined as the minimum of
$E({\bf w}, \lambda)$ with respect to ${\bf w}$.
If $\lambda$ is varied from $\lambda^{(0)}$ to $\lambda^{(0)}+\Delta \lambda$,
the vector ${\bf w}$ which minimizes the energy functional will change
from ${\bf w}^{(0)}$ to
${\bf w}^{(0)} + \Delta {\bf w} $. We can expand the total energy around
${\bf w}^{(0)}$ by a Taylor series:
\begin{equation}
E({\bf w}^{(0)}+\Delta {\bf w}, \lambda^{(0)}+\Delta \lambda) =
 \sum_{p=0}^\infty \sum_{k=0}^\infty {1\over k! p!}
{\delta^{k+p}  E({\bf w}^{(0)}, \lambda^{(0)})
\over \delta {\bf w}^k \delta \lambda^p}
{(\Delta {\bf w})^k} {(\Delta \lambda)^p},
\label{due}
\end{equation}
where we use the notation:
$
{\delta^k E \over \delta {\bf w}^k }(\Delta {\bf w})^k =
\big( \sum_i \Delta w_i {\partial \over \partial w_i} \big)^k E .
$
At a given $\Delta \lambda$ the force is defined by:
\begin{equation}
{\bf f}={\partial E({\bf w}^{(0)}+\Delta {\bf w}, \lambda^{(0)}
+\Delta\lambda) \over
\partial ( {\bf w}^{(0)} + \Delta {\bf w} ) }=
\sum_{p=0}^\infty \sum_{k=1}^\infty {1\over (k-1)! p!}
{\delta^{k+p} E({\bf w}^{(0)}, \lambda^{(0)}) \over
\delta {\bf w}^{k} \delta \lambda^p }
{(\Delta {\bf w})^{k-1}  } {(\Delta \lambda)^p} .
\label{tre}
\end{equation}
The vector $\Delta {\bf w}$ is the solution of the equation obtained from the
extremum condition ${\bf f}=0$.
We now define ${\bf f}^{(n)}$ and $E^{(n)}$ as the force and
the energy to order
$n$ in $\Delta \lambda$. Explicit expressions of these quantities are
obtained by writing $\Delta{\bf w}$ as:
\begin{equation}
\Delta {\bf w} =  {\bf w}^{(1)}  +  {\bf w}^{(2)}  +
\ldots,
\label{qua}
\end{equation}
where ${\bf w}^{(n)}$ is of order $(\Delta \lambda)^n$, and by separating the
various orders in Eq.~(\ref{due}) and Eq.~(\ref{tre}). ${\bf w}^{(n)}$
is obtained from the equation ${\bf f}^{(n)}=0$. Using these definitions
the proof of the $2n+1$ theorem is straightforward.

Since the term quadratic in ${\bf w}^{(l)}$ is of order
$(\Delta \lambda)^{2l}$, $E^{(2n+1)}$ can contain
${\bf w}^{(l)}$ only at linear order if $l>n$.
Under the same condition, we show that the coefficient of ${\bf w}^{(l)}$
in $E^{(2n+1)}$ is zero. To show this it is useful to single out
${\bf w}^{(l)}$
from the product $(\Delta {\bf w})^{k}$ appearing in Eq.~(\ref{due}),
using the relation:
\begin{equation}
(\Delta {\bf w})^k = (\Delta {\bf w}-{\bf w}^{(l)}+ {\bf w}^{(l)})^k=
k {\bf w}^{(l)} (\Delta {\bf w})^{k-1} +
(\Delta {\bf w}-{\bf w}^{(l)})^k+o\big((\Delta\lambda)^{2n+1}
\big),
\label{nuova}
\end{equation}
which is valid for $l>n$. The only term which is linear in ${\bf w}^{(l)}$
is the first term of the r.h.s. of Eq.~(\ref{nuova}).
Inserting this term in Eq.~(\ref{due}) and recalling the
definition of ${\bf f}$, Eq.~(\ref{tre}), we can write
$E^{(2n+1)}$ as:
\begin{equation}
E^{(2n+1)}={\bf w}^{(2n+1)} {\bf f}^{(0)}+\ldots+
{\bf w}^{(l)}{\bf f}^{(2n+1-l)}+
\ldots +{\bf w}^{(n+1)}{\bf f}^{(n)}+ P^{(2n+1)}
({\bf w}^{(1)},\ldots,{\bf w}^{(n)}),
\label{nuova1}
\end{equation}
where $P^{(2n+1)}$ is a polynomial of degree $2n+1$. Since
 ${\bf f}^{(i)}=0$ for every $i$, due to the extremum condition, we obtain
\begin{equation}
E^{(2n+1)}=
 P^{(2n+1)}({\bf w}^{(1)},\ldots,{\bf w}^{(n)}).
\label{teorema}
\end{equation}
This completes the proof of the $2n+1$ theorem.
We note that our formulation does not require any hypothesis on the
specific form of the
energy functional. Therefore our proof can be applied to DFT, to
correlated wavefunctions, and also in contexts other than quantum theory.
Furthermore the present approach combined with a functional with implicit
orthonormalization constraints\cite{Mauri} can be used to derive
the perturbative expansion in cases where the standard approach is
cumbersome,
e.g. the case of DFT when the atoms are described by
Vanderbilt pseudopotentials.\cite{Vanderbilt2}

We now apply the above ideas to the computation of the linear and nonlinear
susceptibilities.
As explained in Ref.~\onlinecite{Vanderbilt} it is possible to
define a total energy functional for a periodic insulating
solid in a finite electric field as:
\begin{equation}
E(|w_0\rangle, F ) = \sum_l \langle w_0 | H +
e F x | w_l \rangle
\bigl ( 2 \delta_{0,l} - \langle w_l | w_0 \rangle \bigr ),
\label{uno}
\end{equation}
where $H$ is the unperturbed Hamiltonian of the solid,
$F$ is the electric field, $x$ is the position operator, $e$ is the
electron charge and
$|w_l\rangle$ are Wannier-like functions
associated to the direct lattice vector $ R_l$,
which are in general non-orthonormal.
The Wannier function $|w_l\rangle$ is obtained by translating
the function centered at the origin by a vector $R_l$.
$|w_0\rangle$ is free to vary
within a real space localization region (LR) of radius $R_c$ centered
at the origin
and it is set equal to zero outside LR.
For simplicity in Eq.~(\ref{uno}) we assume that only the lowest band is
filled, that the system is
one-dimensional and that the total energy describes
independent electrons. Self consistency does not yield any additional
problem.
We stress here that the expectation value of $x$
is well defined for any finite cut-off radius $R_c$.
Furthermore we note that even if no orthogonality constraints are
imposed on the
$|w_l\rangle$, at the minimum they become approximately orthonormal
as shown in Ref.~\onlinecite{Mauri}.

We now recall that the linear and the quadratic susceptibilities
$\chi^{(1)}$ and
$\chi^{(2)}$, are obtained as ${1\over 2} \chi^{(1)} (\Delta F)^2=
 E^{(2)}$ and
${1\over 3} \chi^{(2)}(\Delta F)^3 = E^{(3)}$
where $E^{(n)}$ is the variation of
the energy functional given in Eq.~(\ref{uno}) to order $n$ in the perturbing
field $\Delta F$. From Eq.~(\ref{due}) and Eq.~(\ref{teorema})
with $\Delta \lambda=\Delta F$,
we obtain the expressions:
\begin{equation}
{1\over 2} \chi^{(1)} (\Delta F)^2 ={1\over 2}
{\delta^2 E \over \delta {\bf w}^2}({\bf w}^{(1)})^2
+  {\delta^2 E \over \delta {\bf w} \delta F }
 {\bf w}^{(1)} \Delta  F,
\label{sei}
\end{equation}
\begin{equation}
{1\over 3} \chi^{(2)} (\Delta F)^3 = {1\over 6}{\delta^3 E \over
\delta {\bf w}^3 }
({\bf w}^{(1)})^3+{1 \over 2} {\delta^3 E \over \delta {\bf w}^2 \delta F }
({\bf w}^{(1)})^2 \Delta F,
\label{set}
\end{equation}
where we used the fact that the total energy functional is
linear in the electric field.
The first order variation of the localized orbitals ${\bf w}^{(1)}$ is
obtained either from the condition ${\bf f}^{(1)}=0$ as the solution
of a linear
system:\cite{BGT1}
\begin{equation}
{\delta^2 E \over \delta {\bf w}^2}{\bf w}^{(1)} +
{\delta^2 E \over \delta {\bf w} \delta F } \Delta F=0,
\label{ott}
\end{equation}
or from a direct minimization of $E^{(2)}$, Eq.~(\ref{sei}), with respect
to ${\bf w}^{(1)}$ as in Ref.~\onlinecite{gonze92}.
All the energy derivatives appearing in these
formulas are evaluated at the unperturbed vector ${\bf w}^{(0)}$.

We applied the above results to a 1D model with Hamiltonian
$H=-\nabla^2 + V(x)$ where $V(x)$ is a periodic potential with
period 3, i.e., $V(x+3)= V(x)$.
We chose $V(x)=-\Delta$ if $x\in (-1.5,-0.5] $, $V(x)=\alpha-\Delta$
if $x\in (-0.5,0.5]$, and
$V(x)=0$ if $x\in (0.5,1.5]$.
The parameter $\Delta$ is kept fixed at the value $\Delta=4$ and
$\alpha$ varies between $\alpha = 0$ and $\alpha = \Delta $.
At the two limiting values, the model has inversion symmetry, and therefore
$\chi^{(2)}=0$. Otherwise
the parameter $\alpha$ tunes the value of $\chi^{(2)}$.
We discretized the wavefunctions $w(x)$ on a $N$-point mesh $x_i$ with equal
spacing $\Delta x$. In this representation the action of the Laplacian operator
on the wavefunctions is modelled as a finite difference:
$\nabla^2 w(x_i)=(w(x_{i+1})+
w(x_{i-1}) - 2 w(x_i))/(\Delta x)^2$.
All the calculations are made with $\Delta x = 1/3$.

In Fig.~\ref{f1} we show the $\chi^{(1)}$ values computed from the
analytical derivative of the
total energy, Eq.~(\ref{sei}). These are compared
with the $\chi^{(1)}$ values obtained
from a numerical differentiation of the polarization $P={\delta E \over
\delta F} $ computed at
finite electric fields. The two results are in
perfect agreement. We note that the finite difference calculation
yields the same $\chi^{(1)}$ as  a perturbative approach based on
Bloch wavefunctions.\cite{Vanderbilt}

In the same figure we also show a comparison between the values of
$\chi^{(2)}$, as obtained from the analytical derivative, Eq.~(\ref{set}),
and from a numerical differentiation of
the values of $\chi^{(1)}$ computed at finite electric fields.
Even in this case the two calculations are in perfect agreement.

In Fig.~\ref{f3} we present the values of $\chi^{(2)}$ as a
function of the electric field for $\alpha=2$:
its linearity around the origin allows us to
extract from the slope of the curve the value of
$\chi^{(3)}={2\over 3}{d \chi^{(2)} \over d F}$.

All the above calculations have been done with an $R_c$ value such that the
LR
includes seven unit cells. As shown in Ref.~\onlinecite{Vanderbilt} the
main concern in this type of calculations is the convergence
rate of the studied quantities with respect to the dimensions of the
LR.
In Fig.~\ref{f2} we show how the values of $\chi^{(2)}$ converge as
a function of the size of the LR.

In conclusion we gave a new proof of the $2n+1$ theorem.
We applied it to the total energy of an insulator in a uniform
electric field where the wavefunctions are described by
localized orbitals, with no explicit orthonormalization constraints.
We provided a method to compute analytically the first and
second-order susceptibilities, which is much simpler than
the standard approach.
We tested its accuracy and convergence properties
in a 1D model Hamiltonian.
We believe that the application of this method to DFT and to an
arbitrary kind of perturbation could open the way to
a simple and reproducible computation of high-order derivatives
of the total energy. This will be an efficient way to compute
important properties of real materials such as the Raman tensors
or the nonlinear susceptibilities even in systems with complex unit
cells.

We gratefully acknowledge many usefull discussions with A. Baldereschi,
R. Resta and S. Scandolo. We thank R. Car, A. Canning, P. Fernandez,
and R. Resta for a critical reading
of the manuscript. We acknowledge support by the Swiss National Science
Foundation under Grant No. 21-31144.91.

\begin{figure}
\caption{ Linear (dashed line)
and quadratic (solid line) susceptibilities of the model
system computed analitically with Eqs.~(8),~(9). The results obtained from
numerical differentiation of the polarization (solid squares) and of
the linear susceptibility (open squares), both
computed in finite electric fields, are also shown.}
\label{f1}
\end{figure}

\begin{figure}
\caption{ Quadratic susceptibility as a function of the electric field
for $\alpha=2$.
}
\label{f3}
\end{figure}

\begin{figure}
\caption{ Quadratic susceptibility as a function of the parameter $\alpha$ for
several dimensions of the localization regions.
The curves refers to a localization region equal
to three (long dashed), five (dotted), seven (dashed) and nine (solid line)
unit cells, respectively.}
\label{f2}
\end{figure}

\end{document}